# Structure of a spin ½

B. C. Sanctuary
Department of Chemistry,
McGill University
Montreal Quebec
H3H 1N3 Canada

**Abstract**. The non-hermitian states that lead to separation of the four Bell states are examined. In the absence of interactions, a new quantum state of spin magnitude $1/\sqrt{2}$ is predicted. Properties of these states show that an isolated spin is a resonance state with zero net angular momentum, consistent with a point particle, and each resonance corresponds to a degenerate but well defined structure. By averaging and de-coherence these structures are shown to form ensembles which are consistent with the usual quantum description of a spin.

Keywords: Bell states, Bell's Inequalities, spin theory, quantum theory, statistical interpretation, entanglement, non-hermitian states. PACS: Quantum statistical mechanics, 05.30. Quantum Mechanics, 03.65. Entanglement and quantum non-locality, 03.65.Ud

## 1. INTRODUCTION

In spite of its tremendous success in describing the properties of microscopic systems, the debate over whether quantum mechanics is the most fundamental theory has been going on since its inception[1]. The basic property of quantum mechanics that belies it as the most fundamental theory is its statistical nature[2]. The history is well known: EPR[3] showed that two non-commuting operators are simultaneously elements of physical reality, thereby concluding that quantum mechanics is incomplete, albeit they assumed locality. Bohr[4] replied with complementarity, arguing for completeness; thirty years later Bell[5], questioned the locality assumption[6], and the conclusion drawn that any deeper theory than quantum mechanics must be non-local. In subsequent years the idea that entangled states can persist to space-like separations became well accepted[7].

However the EPR coincidence experiments on entangled photons[8,9] that demonstrate violation of Bell's Inequalities cannot be rationalized within quantum mechanics but are purported to depend on non-local connectivity between the pair[10]. Non-locality is a concept, as yet, not understood[11].

If changes are made to quantum theory which can explain notions like non-locality, then these must simply resolve quantum theory into something deeper and restore it again in an appropriate limit. Replacing such a successful theory is not the goal. In this paper the treatment involves spin ½ only. It has been shown that the Bell states[12,13], maximally entangled within quantum theory, separate into products of single particle states when a specific non-hermiticity is introduced. Motivated by this property, these non-hermitian states are shown to be consistent with both EPR's requirement of reality[3] and, after ensemble averaging, also quantum theory, which suppresses or treats the information associated with the non-hermiticity as random.

The consequences of the changes introduced here are mostly benign. The non-hermitian state operators retain trace 1 and are idempotent, consistent with pure states. They have real eigenvalues, which are the same as those when the non-hermiticity is dropped. The only mathematical difference is the two eigenstates are non-orthogonal.

The most significant change is in the interpretation of the non-hermitian states. Rather than being the statistical operators of quantum theory, they are treated here as representing the pure spin states of a single particle. This is based upon the premise that single particles exist and therefore there must be a way of representing their states. Silver atoms, oxygen molecules, electrons *etc*. are the individual components that make up systems containing enough particles for reliable and reproducible measurements. In analogy with the ideas of classical statistical mechanics, the ensembles that form the quantum states are proposed to be composed of individual particles whose states are non-hermitian.

In the following it is first argued that the non-hermitian spin states suggest an underlying physical structure for a spin ½. Using common requirements of quantum theory, the states of an isolated spin are found to be 8-fold degenerate with eigenvalues of ±1. The view that emerges is one of resonance between these degenerate spin states that is not inconsistent with the quantum conclusion that an isolated spin acts like a point particle[14]. When such spins are influenced by local or external magnetic fields, however, all states randomize except those quantized along the field axis, leading to the usual view of a spin ½ as, for example, filtered in a Stern-Gerlach experiment. In addition, it is shown that phase randomization of an ensemble of non-hermitian spins leads to the usual hermitian spin states of quantum theory. In fact one conclusion is the non-hermiticity describes a phase, and only in unique situations (*e.g.* apparently to date only coincidence experiments on photons) can it be detected.

## 2. NON-HERMITIAN STATES

A singlet state is one of the entangled Bell state,

$$|\psi_{12}^-\rangle = \frac{1}{\sqrt{2}}\left(|+\rangle_Z^1|-\rangle_Z^2 - |-\rangle_Z^1|+\rangle_Z^2\right) \quad (2.1)$$

where the $|\pm\rangle_Z^i$ are the usual eigenstates[15] of the Pauli spin operator $\sigma_z$ with eigenvalues of $\pm 1$. Since the singlet is isotropic, the choice of the quantization axis, $\hat{\mathbf{Z}}$ (subscript Z), is arbitrary as long as it coincides for both spins. For this reason the LHS of Eq.(2.1) has no dependence on $\hat{\mathbf{Z}}$. The statistical, or density, operator[16,17] for this two particle spin state is given by

$$\rho_{\psi_{12}^-} = |\psi_{12}^-\rangle\langle\psi_{12}^-| \quad (2.2)$$

and displays quantum interference between the off-diagonal elements $|\pm\rangle_Z^1|\mp\rangle_Z^2$ and $|\mp\rangle_Z^1|\pm\rangle_Z^2$. Such terms are responsible for its entanglement[18]. There is no subscript Z on Eq.(2.2) because its form is invariant to the choice of coordinate system in three dimensional physical space. Any coordinate frame will do. If, for example, the spins are filtered, then the quantization axis is in the direction of the external magnetic field, $\hat{\mathbf{Z}}$. Hence $\hat{\mathbf{R}} = \{\hat{\mathbf{X}}, \hat{\mathbf{Y}}, \hat{\mathbf{Z}}\}$ represents a laboratory coordinate frame of the Stern-Gerlach filter.

In contrast, since any axis of quantization is acceptable for the singlet state, and in the absence of any magnetic interactions, we treat the coordinate frame, $\hat{\mathbf{r}} = \{\hat{\mathbf{x}}, \hat{\mathbf{y}}, \hat{\mathbf{z}}\}$, as arbitrary. The laboratory frame is only one special case and all are related by rotations. A difference in approach here lies between the statistical operator for the pure states and the state operator for a single spin. The former can be represented in the laboratory frame of the filter,

$$\rho_Z = \frac{1}{2}(I + n_Z\sigma_Z) \quad (2.3)$$

where $n_Z = \pm 1$. In contrast the pure state operator that describe the states of a single spin is proposed to have the fundamental form which is valid in any coordinate frame in 3D space, $\hat{\mathbf{r}} = \{\hat{\mathbf{x}}, \hat{\mathbf{y}}, \hat{\mathbf{z}}\}$,

$$s_z(n_z n_x n_y) \equiv \frac{1}{2}(I + n_z\sigma_z + n_x\sigma_x + in_y\sigma_y) \quad (2.4)$$

where $(n_z, n_x, n_y) = \{\pm 1, \pm 1, \pm 1\}$, with eight possible permutations. This non-hermitian form, Eq.(2.4), is proposed because it leads to separation of Bell states as for example the singlet[12]

$$s_{\Psi_{12}^-} = \frac{1}{8}\begin{bmatrix}\left[s_z^1(+1,+1,+1)\otimes s_z^2(-1,-1,+1) + s_z^1(+1,+1,+1)^\dagger \otimes s_z^2(-1,-1,+1)^\dagger\right] + \\ \left[s_z^1(-1,-1,+1)\otimes s_z^2(+1,+1,+1) + s_z^1(-1,-1,+1)^\dagger \otimes s_z^2(+1,+1,+1)^\dagger\right] + \\ \left[s_z^1(+1,-1,+1)\otimes s_z^2(-1,+1,+1) + s_z^1(+1,-1,+1)^\dagger \otimes s_z^2(-1,+1,+1)^\dagger\right] + \\ \left[s_z^1(-1,+1,+1)\otimes s_z^2(+1,-1,+1) + s_z^1(-1,+1,+1)^\dagger \otimes s_z^2(+1,-1,+1)^\dagger\right]\end{bmatrix} \quad (2.5)$$

where † denotes the hermitian conjugate operator. That is, mathematically at least, the statistical operator, Eq.(2.2) and the state for one pair of spins in a singlet, Eq.(2.5), are identical. In essence because the singlet state is isotropic, any pair of spins with arbitrary common frame, $\hat{\mathbf{r}} = \{\hat{\mathbf{x}}, \hat{\mathbf{y}}, \hat{\mathbf{z}}\}$, contributes equally and identically to the ensemble that the statistical operator, Eq.(2.2), represents. In Eq.(2.5) the symbol $\otimes$ denotes the direct product between the two spin states.

Hermitian statistical operators are related to their state vectors in diagonal form $\rho = |\psi\rangle\langle\psi|$, see *e.g.,* Eq.(2.2). In contrast the non-hermitain states can only be written in off-diagonal form $s = |\varphi\rangle\langle\varphi'|$. Therefore it is necessary to treat the non-hermtitian states as operators in Hilbert-Schmidt space with Hilbert-Schmidt norm[19]. The time dependence is then given by the quantum Liouville equation rather than the Schrödinger equation.

The measurement postulate remains the same as $\langle \hat{A} \rangle = \text{Tr}[\rho \hat{A}]$. The relationship between the non-hermitian states and the density operator involves ensemble averaging. In the case of the singlet, its isotropy renders it invariant in 3D space and therefore the average over all different orientations, $\hat{\mathbf{r}}$, gives,

$$\rho_{\psi_{\bar{12}}} = \overline{s_{\psi_{\bar{12}}}}^{EA} = \int p(\hat{\mathbf{r}}) s_{\psi_{\bar{12}}} d\hat{\mathbf{r}} = s_{\psi_{\bar{12}}} \tag{2.6}$$

where the bar denotes ensemble averaging (EA) and $p(\hat{\mathbf{r}})$ is a distribution function. Further discussion of the Bell states is deferred to a later publication[13]. In the following, the non-hermitian states that separate the Bell states are investigated.

## 3. PROPERTIES OF THE NON-HERMITIAN STATES

As prepared for experiment, and after many particles have passed a Stern-Gerlach filter, the eigenvalues of $\sigma_Z$ are +1 and -1 with orthogonal quantum states in the laboratory frame of

$$|+\rangle_Z = \begin{pmatrix} 1 \\ 0 \end{pmatrix}_Z \text{ for } +1 \text{ and } |-\rangle_Z = \begin{pmatrix} 0 \\ 1 \end{pmatrix}_Z \text{ for } -1 \tag{3.1}$$

Although from a random source on the average, fifty percent of the spins are deflected up and fifty percent down, there is no possibility within quantum theory to predict a priori which way a spin will be deflected, or which slit an electron passes in a double slit experiment. These are examples of the statistical nature of quantum theory.

In contrast the non-hermitian states of the single events before reaching the Stern-Gerlach filters are deterministic by the definition Eq.(2.4), but this is usually masked by the degeneracy associated with them. There are eight different choices of integers in Eq.(2.4) each giving the same eigenvalues and displaying states that differ from the others only in sign. Whereas the eigenvalues of $\sigma_Z$

$$\sigma_Z = \begin{pmatrix} 1 & 0 \\ 0 & -1 \end{pmatrix} \tag{3.2}$$

are ±1 with eigenstates give by Eq.(3.1), the non-hermitian spin operator in Eq.(2.4) have eight different forms due to the permutation of integers $(n_z, n_x, n_y)$. Each choice corresponds to one of the octants in the frame, $\hat{\mathbf{x}}, \hat{\mathbf{y}}, \hat{\mathbf{z}}$. One of these is

$$\sigma_z + \sigma_x + i\sigma_y = \begin{pmatrix} 1 & 2 \\ 0 & -1 \end{pmatrix}_z \tag{3.3}$$

and has the same eigenvalues of ±1 but with non-orthogonal eigenstates,

$$|+\rangle_z = \begin{pmatrix} 1 \\ 0 \end{pmatrix}_z \text{ for } +1 \text{ and } |-\rangle_x = \frac{1}{\sqrt{2}} \begin{pmatrix} 1 \\ -1 \end{pmatrix}_x \text{ for } -1 \tag{3.4}$$

One eigenvalue is associated with the $\hat{\mathbf{z}}$ axis and the other with the $\hat{\mathbf{x}}$ axis. For all choices of the integers in Eq.(2.4), the eigenvalues are always +1 and -1 and the associated eigenstates pairs for each octant are the various non-orthogonal combinations of

$$\begin{pmatrix} 1 \\ 0 \end{pmatrix}_z ; \begin{pmatrix} 0 \\ 1 \end{pmatrix}_z ; \frac{1}{\sqrt{2}} \begin{pmatrix} 1 \\ -1 \end{pmatrix}_z ; \frac{1}{\sqrt{2}} \begin{pmatrix} 1 \\ 1 \end{pmatrix}_z . \tag{3.5}$$

That is the eigenvalues of ±1 are eight-fold degenerate. There are no eigenstates associated with *y* component.

The view that emerges is consistent with the way particles with structure are treated in, for example, spectroscopy. Consider a symmetric top molecule which has internal angular momentum which can be observed spectroscopically from the laboratory frame. The usual technique is to define a body fixed coordinate frame, $\hat{\mathbf{x}}, \hat{\mathbf{y}}, \hat{\mathbf{z}}$, and express the internal properties in terms of, in the case of rotations, angular momentum around two axes due to it having two different moments of inertia. These are generally called the rotational angular momentum *J* and the projection of this on its symmetry axis, *K*. Once these are defined, an ensemble of such particles, all with different body frames, is transformed to the laboratory frame where they are observed statistically.

At any instant, a spin can exist in only one of its eight octants, but all are assumed to be equally probable in the absence of interactions. That is, whereas a single spin is oriented and well defined in one octant, it is impossible, due to degeneracy, to distinguish it from others that are oriented in different octants of the same body frame, $\{\hat{\mathbf{x}},\hat{\mathbf{y}},\hat{\mathbf{z}}\}$.

The structure of a spin, as seen from Eq.(2.4), displays two angular momentum, one associated with the body frame $\hat{\mathbf{z}}$ axis, and the other with the $\hat{\mathbf{x}}$ axis, see also Eq.(3.4). This leads to a two dimension structure for each of the eight orientations within the body frame, see Figure 1. The non-hermiticity is a result of the term $i\sigma_y$ which is not interpreted here as angular momentum but rather as a rotation operator. It orients the 2D spin in 3D space via the quantum equivalent of the cross products, $\sigma_z \sigma_x = i\sigma_y$, see Figure 1.

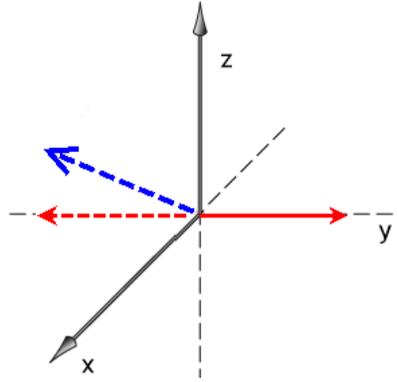

**Figure 1.** A 2D spin is oriented in one octant of its body frame, $\{\hat{\mathbf{x}},\hat{\mathbf{y}},\hat{\mathbf{z}}\}$ in 3D real space. The two angular momenta of Eq.(2.4) are directed along the $\hat{\mathbf{z}}$ and $\hat{\mathbf{x}}$ axes. The components along $\hat{\mathbf{y}}$ and the $-\hat{\mathbf{y}}$ depict two identical views if the $\hat{\mathbf{z}}$ and $\hat{\mathbf{x}}$ axes are indistinguishable. The vector that bisects the $\hat{\mathbf{z}}\hat{\mathbf{x}}$ plane is the resonance spin state of length √2.. The spin can also be oriented in the other quadrants in the $\hat{\mathbf{z}}\hat{\mathbf{x}}$ plane, and all four are treated as resonance structures.

## The hermitian resonance states of a single spin ½.

A basic tenet that emerges from the development here is the non-hermiticity, needed for its physical reality of spin and the separability of entangled states, gives way to hermitian states except in special cases. One such case is seen from Eq.(2.5) where the singlet is composed of pairs of non-hermitian states even though these combine to give the singlet, which is hermitian. Likewise, as shown in below, when in the presence of a magnetic interaction, the non-hermiticity is phase randomized. In this subsection it is shown, due to the quantum requirement of indistinguishablility, the states for an isolated spin are hermitian.

Analogy can be made with the description of chemical bonds, such as the hydrogen molecule. In the early treatments, three contributions to the bond strength were identified being the overlap integral, the Coulomb integral and the exchange integral. The last arises due to the indistinguishablility of the two arrangements, $H_A$-$H_B$ and $H_B$-$H_A$ and this purely quantum resonance or exchange component accounts for 90% of the hydrogen molecule's bond strength.

If the two orthogonal components expressed by the state Eq.(2.4) predict spin to be a two dimensional object, then in the absence of external forces and in its body frame, the two axes of quantization, $\hat{\mathbf{x}}$ and $\hat{\mathbf{z}}$ are indistinguishable. In order to ensure this property, the right $\hat{\mathbf{x}},\hat{\mathbf{y}},\hat{\mathbf{z}}$ and left $\hat{\mathbf{x}},-\hat{\mathbf{y}},\hat{\mathbf{z}}$ handed body frames are equivalent, see Firue 1, and this leads to the hermitian state of a single spin,

$$s_{z,\sqrt{2}}\left(\hat{\mathbf{n}}_{n_z n_x}\right) = \frac{1}{2}\left(s_z\left(n_z n_x n_y\right) + s_z\left(n_z n_x n_y\right)^\dagger\right) = \frac{1}{2}\left(I + \sqrt{2}\boldsymbol{\sigma}\cdot\hat{\mathbf{n}}_{n_z n_x}\right) \qquad (3.6)$$

The unit vectors in Eq.(3.6) bisects the four indistinguishable axes of the body frame $(n_z, n_x) = \{\pm 1, \pm 1\}$,

$$\hat{\mathbf{n}}_{n_z n_x} = \frac{1}{\sqrt{2}} (n_z \hat{\mathbf{z}} + n_x \hat{\mathbf{x}}) \tag{3.7}$$

while the anti-hermitian difference is given by a rotation operator<sup>Error! Bookmark not defined.</sup>,

$$\frac{1}{2} \left( s_z(n_z n_x n_y) - s_z(n_z n_x n_y)^\dagger \right) = \frac{1}{2} i n_y \sigma_y = \pm \frac{i \sigma_y}{2} \tag{3.8}$$

Within this theory equation (3.6) represents a resonance state formed from the superposition of the indistinguishable quantization axes with states given by, for example, Eq.(3.4). Whereas the spin states described by the statistical operator, Eq.(2.3), have eigenvalues of $\pm 1$, those of Eq.(3.6) are $\pm\sqrt{2}$. Depending of the values of $n_z n_x n_y$ four sets of orthogonal eigenstates are found. For $\pm(\sigma_z + \sigma_x)$ two sets are,

$$\pm \begin{pmatrix} 1 & 1 \\ 1 & -1 \end{pmatrix}_z \text{ gives eigenvalues of } \pm\sqrt{2} \quad \text{and eigenstates of} \quad |\pm\rangle^{\sqrt{2}}_{\pm 1, \pm 1, z} = \frac{\pm 1}{\sqrt{2(2+\sqrt{2})}} \begin{pmatrix} 1+\sqrt{2} \\ 1 \end{pmatrix}_z \tag{3.9}$$

$$|\mp\rangle^{\sqrt{2}}_{\pm 1, \pm 1, z} = \frac{\pm 1}{\sqrt{2(2+\sqrt{2})}} \begin{pmatrix} 1 \\ -1-\sqrt{2} \end{pmatrix}_z$$

and from $\pm(\sigma_z - \sigma_x)$ the two sets are,

$$\pm \begin{pmatrix} 1 & -1 \\ -1 & -1 \end{pmatrix}_z \text{ gives eigenvalues of } \pm\sqrt{2} \quad \text{and eigenstates of} \quad |\pm\rangle^{\sqrt{2}}_{\pm, 1\mp, 1, z} = \frac{\pm 1}{\sqrt{2(2+\sqrt{2})}} \begin{pmatrix} -1-\sqrt{2} \\ 1 \end{pmatrix}_z \tag{3.10}$$

$$|\mp\rangle^{\sqrt{2}}_{\pm, 1\mp, 1, z} = \frac{\pm 1}{\sqrt{2(2+\sqrt{2})}} \begin{pmatrix} 1 \\ 1+\sqrt{2} \end{pmatrix}_z$$

Geometrically these eigenstates lie at angles of 45º between the four sets of the two axes defined by $n_z \hat{\mathbf{z}}$ and $n_x \hat{\mathbf{x}}$, Eq.(3.7), and are a superposition of the non-orthogonal states in Eq.(3.4). The resulting spin of magnitude $\sqrt{2}$ is a pure quantum resonance in origin. In terms of the usual spin magnitude $\hbar/2$, the operator for the above spin states for a single particle is defined by,

$$\mathbf{S}_{\sqrt{2}} \equiv \frac{\hbar}{\sqrt{2}} \boldsymbol{\sigma} \tag{3.11}$$

If the environment is not isotropic, such as when the spin interacts with local or external magnetic fields, the two axes become distinguishable and the $\sqrt{2}$ resonance state cannot be formed. The spin, Eq.(3.11) is predicted to exist in an isotropic environment only, in the absence of all interactions.

The following three subsections show that the non-hermitian states give a consistent description of isolated spins; spins in a bulk isotropic system; and in the presence of an external magnetic field. In all cases, the ensemble averaging and de-coherence renders to states hermitian.

## An isolated spin

In the previous section the resonance state can exists only for an isolated spin oriented in any one of the quadrants of its body frame defined by the $\hat{\mathbf{x}}, \hat{\mathbf{z}}$ plane, see Figure 1. However there is no discernable difference between any of these four orientations so that the state of an isolated spin is given as an equal superposition of the four,

$$s_{z,\sqrt{2}} = \frac{1}{4} \left( s_{z,\sqrt{2}}(\hat{\mathbf{n}}_{11}) + s_{z,\sqrt{2}}(\hat{\mathbf{n}}_{-1-1}) + s_{z,\sqrt{2}}(\hat{\mathbf{n}}_{1-1}) + s_{z,\sqrt{2}}(\hat{\mathbf{n}}_{-11}) \right) = \frac{I}{2} \tag{3.12}$$

The view here is different from the classical meaning of objective reality. Here a single spin is viewed as being in a well defined state described by Eq.(3.6) at an instant of time, and those states are dispersion free. However equally probable is its orientation in the three other quadrants. Since the sum of the four resonances gives the identity, and a state of zero net angular momentum, the point particle description of intrinsic angular momentum is corroborated.

Resonance is a common property of quantum systems. By definition, resonance means that there are a number of equivalent structures of a system and at any instant only one of these exists. Common chemical techniques on molecules make use of resonance to promote the reactivity of one structure over the others. Likewise, the resonance state for the 2D spin has the same meaning. At any instant the spin can be in any one of its eight degenerate states. This is consistent with the inability of quantum theory to predict with certainty which resonance state a spin actually occupies. In the vast majority of cases such information is irrelevant and so, in deference to its success, is lost in the statistical nature of quantum theory.

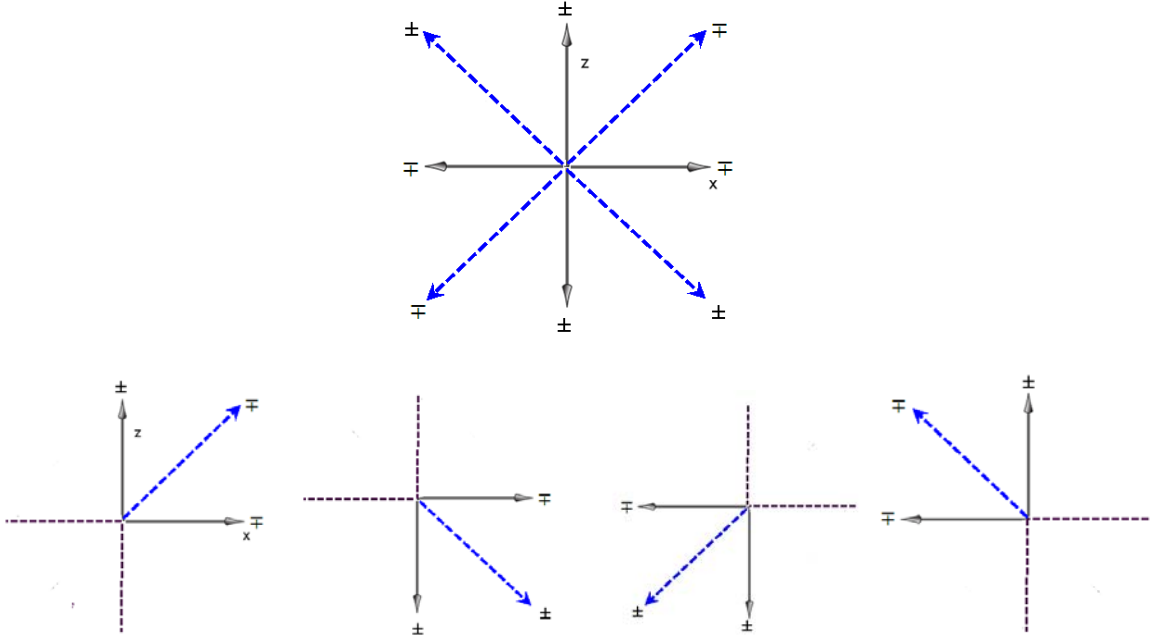

**Figure 2**. Top, a representation of the superposed resonance states of a spin ½ obtained from Eq.(3.12) which sum to the identity consistent with a point particle. Bottom, the four resonance structures separated with each representing two degenerate dispersion-free states. The plus-minus values represent the signs of the angular momenta.

## De-coherence in the laboratory frame.

In this subsection the spins oriented in their body frame are transformed to the laboratory frame where it is shown that de-coherence renders the macroscopic states hermitian and leads to consistency with the usual results from quantum theory.

The laboratory frame, $\hat{\mathbf{X}}, \hat{\mathbf{Y}}, \hat{\mathbf{Z}}$, is chosen randomly in any convenient direction in isotropic 3D space. The body and the laboratory frames differ by a rotation by $\theta, \phi$

$$\hat{\mathbf{z}} = \cos\theta \hat{\mathbf{Z}} + \sin\theta \cos\varphi \hat{\mathbf{X}} + \sin\theta \sin\varphi \hat{\mathbf{Y}}$$
$$\hat{\mathbf{x}} = -\sin\theta \hat{\mathbf{Z}} + \cos\theta \cos\varphi \hat{\mathbf{X}} + \cos\theta \sin\varphi \hat{\mathbf{Y}} \qquad (3.13)$$
$$\hat{\mathbf{y}} = -\sin\varphi \hat{\mathbf{X}} + \cos\varphi \hat{\mathbf{Y}}$$

The states in the body and laboratory frames are related by

$$|+\rangle_z = \begin{pmatrix} \cos\theta/2\, e^{-i\varphi/2} \\ \sin\theta/2\, e^{i\varphi/2} \end{pmatrix}_Z \qquad |-\rangle_x = \frac{1}{\sqrt{2}} \begin{pmatrix} -(\cos(\theta/2)+\sin(\theta/2))e^{-i\varphi/2} \\ (\cos(\theta/2)-\sin(\theta/2))e^{+i\varphi/2} \end{pmatrix}_Z$$

$$|-\rangle_z = \begin{pmatrix} -\sin(\theta/2)e^{-i\varphi/2} \\ \cos(\theta/2)e^{+i\varphi/2} \end{pmatrix}_Z \qquad |+\rangle_x = \frac{1}{\sqrt{2}} \begin{pmatrix} (\cos(\theta/2)-\sin(\theta/2))e^{-i\varphi/2} \\ (\cos(\theta/2)+\sin(\theta/2))e^{i\varphi/2} \end{pmatrix}_Z$$

(3.14)

and the Pauli spin components between the two frames take the form

$$\sigma_z = \cos\theta\sigma_Z + \frac{1}{2}\sin\theta\left(e^{-i\phi}\sigma_+ + e^{+i\phi}\sigma_-\right)$$

$$\sigma_x = -\sin\theta\sigma_Z + \frac{1}{2}\cos\theta\left(e^{-i\phi}\sigma_+ + e^{+i\phi}\sigma_-\right) \quad (3.15)$$

$$i\sigma_y = +\frac{1}{2}\left(e^{-i\phi}\sigma_+ - e^{+i\phi}\sigma_-\right)$$

which are expressed in terms of the raising and lowering operators $\sigma_\pm = \sigma_X \pm i\sigma_Y$. Since transformation from any body frame octant to the laboratory frame differs only by signs, the treatment is restricted to the first octant: the others follow similarly. Substitution of Eqs.(3.15) into Eq.(2.4) gives

$$s_z(111) \equiv \frac{1}{2}\left(I + \cos\theta\sigma_Z - \sin\theta\sigma_Z + \frac{1}{2}\left(e^{-i\phi}\sigma_+(1+\sin\theta+\cos\theta) - e^{+i\phi}\sigma_-(1-\sin\theta-\cos\theta)\right)\right) \quad (3.16)$$

This represents the pure state of a single randomly oriented non-hermitian spin in the laboratory frame where the phase coherence, via $\phi$, is displayed. Random fluctuations cause phases to undergo de-coherence[20], $\overline{e^{\pm i\phi}}^D = 0$ indicated by the overbar. This or integration over a distribution of many spins with random phase angles, leads directly to the hermitian statistical operator,

$$\rho_Z^D \equiv \overline{s_z(111)}^D = \frac{1}{2}\left(I + (\cos\theta - \sin\theta)\sigma_Z\right) \quad (3.17)$$

which is generally a mixed state with the trace of the square given by

$$\text{Tr}\left[\rho_Z^D\right]^2 = 1 - \cos\theta\sin\theta \quad (3.18)$$

If we consider a beam of particles with spin moving in the laboratory **Y** direction, then de-coherence shows that there is no spin component in that direction, and the expectation values will be real. Pure state are found in cases: $\theta = 0, \pi/2$. The case $\theta = 0$ corresponds to an ensemble of spins with random phases all with the same body frame $\hat{z}$ axes parallel or antiparallel with the laboratory axis ($\hat{\mathbf{Z}} = \pm\hat{z}$). The eigenstates are given by $|\pm\rangle_Z$, see Eq.(2.1). The second case $\theta = \pi/2$ has opposite eigenvalues to the other pure case and corresponds to $\hat{\mathbf{Z}} = \pm\hat{x}$. Since it is impossible to distinguish one body axis from the other, the states are the same, but opposite, respectively $|\mp\rangle_Z$. In the above development, the laboratory frame can be oriented in any direction because the space is assumed to be isotropic. Finally for mixed states, $\theta \neq 0, \pi/2$, averaging over a random distribution of $\theta$ leads to the statistical operator for an unpolarized mixture,

$$\rho \equiv \overline{s_{x,y,z}}^{EA} = \frac{I}{2} \quad (3.19)$$

The presence of a magnetic field destroys the isotropy and is treated next.

## Non-hermitian spins in external fields.

An external field with major axis in the $\hat{\mathbf{Z}}$ direction destroys the isotropy of real space and causes the two body axes of each 2D spin to precess differently. Over an ensemble, all states randomize except those where one of the two body axes, $\hat{x}$ or $\hat{z}$, aligns with $\hat{\mathbf{Z}}$. In this calculation an unpaired electron spin in an atom is used that can interact with either local moments or external magnetic fields. If a Stern-Gerlach filter is used, the magnetic field must be inhomogeneous in order to provide a coupling between the linear momentum and the spin angular momentum[21]. The objective here is not to show that beams are deflected up and down, but rather that one of the components aligns with the field while the others are randomized by ensemble averaging due to their time dependence in the field. A homogeneous field therefore suffices.

Although a completely isolated spin forms the √2 spin, this is destroyed as it approaches a magnetic field. Therefore the non-hermitian state is used in this calculation. Assume the external field is in the laboratory $\hat{\mathbf{Z}}$ direction, given by $\mathbf{H}_o = H_o\hat{\mathbf{Z}}$. Therefore the hamiltonian for the Zeeman interaction is,

$$H = -\boldsymbol{\mu}_e \cdot \hat{\mathbf{Z}}H_o \quad (3.20)$$

in terms of the magnetic moment of a spin, $\mu_e$ and the filed strength is $H_o$. Due to a single spin having two quantization axes, it is assumed a magnetic moment is associate with each, given in terms of the Bohr magneton, $\mu_\beta$, and the $g$-factor along each body frame axis,

$$\mu_{ez} = -g\mu_\beta \sigma_z \hat{\mathbf{z}} \text{ and } \mu_{ex} = -g\mu_\beta \sigma_x \hat{\mathbf{x}} \tag{3.21}$$

There is no dipole-dipole interaction within a single electron because of the orthogonality of $\hat{\mathbf{x}}$ and $\hat{\mathbf{z}}$, so the magnetic moment is given by

$$\mu_e = \mu_{ez} - \mu_{ex} = -g\mu_\beta \left(\sigma_z \hat{\mathbf{z}} - \sigma_x \hat{\mathbf{x}}\right) \tag{3.22}$$

Using the quantum Liouville equation,

$$i\hbar \frac{d\rho}{dt} = [H, \rho] \tag{3.23}$$

with the body frame state operator, Eq.(2.4) and the hamiltonian, Eq.(3.20) leads to

$$\frac{d(\mu_{ez} - \mu_{ex})}{dt} = -i\omega_o \left\{(\mu_{ex} \cos\theta + \mu_{ez} \sin\theta)\right\} \tag{3.24}$$

with $\omega_o \equiv g\mu_\beta H_o$ and is independent of the phase angle $\phi$. This separates into two equations with solutions,

$$\frac{d\mu_{ex}}{dt} = +i\cos\theta\omega_o \mu_{ex} \to \mu_{ex}(t) = e^{i\cos\theta\omega_o t} \mu_{ex}(0)$$

$$\frac{d\mu_{ez}}{dt} = -i\sin\theta\omega_o \mu_{ez} \to \mu_{ez}(t) = e^{-i\sin\theta\omega_o t} \mu_{ez}(0) \tag{3.25}$$

showing that each body axis of quantization accumulates phase in opposite directions and with a frequency that is modulated by the angle the magnetic moments make with the external magnetic field.

This calculation shows that all states are averaged to zero except the component aligned with the field, hence for the states, $\theta = 0, \pi/2$ one or the other body axis is aligned with the field axis. For $\theta = 0$

$$\overline{\mu_{ez}(t)}^{EA} = \mu_{eZ} \qquad \overline{\mu_{ex}(t)}^{EA} = 0 \tag{3.26}$$

or for $\theta = \pi/2$

$$\overline{\mu_{ez}(t)}^{EA} = 0 \qquad \overline{\mu_{ex}(t)}^{EA} = \mu_{eZ} \tag{3.27}$$

The magnetic moment is defined in the laboratory axis, $\mu_{eZ} = -g\mu_\beta \sigma_z \hat{\mathbf{Z}}$ in accord with the usual results from quantum theory and the pure states are given by Eq.(3.1), with a single laboratory axis of quantization.

## 4. DISCUSSION.

This work was motivated by the discovery that the four Bell states are separable when various permutations of the non-hermitian spin state, Eq.(2.4), are used. The usual concern regarding non-hermititian states, *i.e.* they lead to complex expectation values, is not founded here. The state operators have the same real eigenvalues; have unit trace; are idempotent; and differ by having non-orthogonal eigenstates. Moreover, in the cases considered, the non-hermiticity gives way to the usual hermitian states of quantum theory. Isolated spins, due to indistinguishablility of the two axes, lead to hermitian resonance states of magnitude √2. The non-hermiticity is also suppressed when representing Bell states as separable, Eq.(2.5), even though the non-hermiticity of the two spins in necessary for the separability. In three cases treated, the non-hermititian states of the single particles form hermitian ensembles for: isolated spins; spins in the laboratory frame; and spins in a magnetic field, all of which are consistent with the quantum view of a spin.

From Eq.(3.16) the rotation operator, $i\sigma_y$ is a phase with the interpretation of orienting the 2D spin in 3D space. It is determined from the values of the two angular momentum, $\sigma_z \sigma_x = i\sigma_y$. The question arises as to the significance of knowing the orientation of single spins in 3D space. That knowledge is obfuscated by other resonances and degeneracy making it virtually unfeasible to distinguish one orientation from another in the same body frame. A consequence of accepting that spins have a specific well defined structure on the one hand satisfies the requirement of objective reality of the 2D spin, and on the other gives a basis for the statistical nature of quantum theory. Within quantum theory, and by the hermitian postulate, knowledge associated with the phase is lost giving rise to entanglement and dispersion. This is the price paid for retaining hermitian ensemble averages over the non-hermitian states, but is also a strength of quantum theory which suppresses information which is usually random.

The notion of objective reality is different here from the classical definition. If it were possible to perform reproducible experiments on a single particle which did not disturb it, then the three operators describing a spin ½, $\{\sigma_z, \sigma_x, i\sigma\}$ must all exist simultaneously. Quantum effects; resonances; and de-coherence at the quantum level can randomize the information due to the non-hermiticity.

Extension to existing theories should lead to new predictions. The major prediction here is the existence of the √2 spin. The only experiments known to date which are taken as support for non-local connectivity through violation of Bell's Inequalities, are the various coincidence experiments done with photons. Although quantum theory correctly predicts the correlation between entangled pairs, it does not give a basis for understanding why various filter settings lead to those violations. In a later publication it is shown that the presence of the √2 spin is responsible for such settings, and can also rationalize the violation of the CHSH from of Bell's Inequalities by 2√2.

It remains to be shown that the other properties of spin are consistent with the 2D structure presented here. These include evaluation of the electron *g*-factor and application to experiments such as the double slit on electrons which is expected to be dependant on the resonance states of a single spin.

## ACKNOWLEDGMENT.

This work is supported by a Discovery Grant from the Natural Sciences and Engineering Research Council of Canada (NSERC).